\documentclass[12pt,reqno]{amsart}
\usepackage[cp1251]{inputenc}
\usepackage[english]{babel}
\usepackage{amsmath,amssymb,amsfonts,cite}

\newtheorem{theorem}{Theorem}

\numberwithin{equation}{section}

\let\p\partial

\def\phi{\varphi}
\let\bsy\boldsymbol
\let\ge\geqslant
\let\le\leqslant
\let\t\tilde

\newcommand{\I}{\text{Im}\,}

\let\ds\displaystyle

\def\ba{\begin{aligned}} 
\def\ea{\end{aligned}}

\def\n{\nonumber}

\newcounter{rem}
\newcommand{\rem}{\addtocounter{rem}{1}\textbf{Remark \therem.} }

\numberwithin{equation}{section}

\begin{document}

\thispagestyle{empty}

\baselineskip=7mm
\thispagestyle{empty}
\begin{center}
{\Large\bf Integrable Hamiltonian equations  of fifth order with the Hamiltonian operator $\boldsymbol D_x$}
\end{center}

 \vskip5mm \hfill
\begin{minipage}{12.5cm}
\baselineskip=15pt
{\bf A.G.  Meshkov ${}^{1}$
 and
   V.V. Sokolov ${}^{2}$} \\ [1ex]
{\footnotesize
${}^1$ University -- UNPK, Orel, Russia 
 \\
${}^{2}$ Landau Institute for Theoretical Physics, Moscow, Russia }\\

\end{minipage}
\begin{center}
\begin{minipage}[c]{140mm}
\small ABSTRACT. All non-equivalent integrable evolution   equations of the  fifth  order of the form $u_t=D_x\frac{\delta H}{\delta u}$  are found. 
\end{minipage}
\end{center}

\section{Introduction.}
Consider integrable Hamiltonian evolution equations of the form 
\begin{equation}\label{Hamgen}
u_t=D_x\left(\frac{\delta H}{\delta u}  \right).
\end{equation}
Here $H(x,u,u_x,...)$ is the Hamiltonian and $D_x$ is the total $x$-derivative.  By integrability we mean the existence of infinite hierarchy of 
commuting flows (or the same higher symmetries) for (\ref{Hamgen}). 
The celebrated KdV equation with $H=-\frac{1}{2}\,u_x^2+u^3$  provides the simplest example of such an equation. 

In \cite{meshsok} integrable equations (\ref{Hamgen}) of the third order have been investigated. Such equations have the form 
\begin{equation}\label{Ham}
u_t=D_x\left(\frac{\p H}{\p u}-D_x\frac{\p H}{\p u_x}\right),
\end{equation}
where $H=H(x,u,u_x).$ It follows from the simplest necessary integrability condition \cite{ss,mss} that there exist two different kinds of integrable Hamiltonians:
$$
H=a_1 u_x^2+a_2 u_x+a_3 \qquad {\rm and} \qquad H=k_1 u_x+k_2+\sqrt{k_3 u_x^2+k_4 u_x+k_5}, 
$$
where $a_i$ and $k_i$ are some functions of $u$ and $x$. Since the function $H$ is defined up to the equivalence
$H\to H+D_x f(x,u)+\lambda u,$  we may put without loss of generality $a_2=k_1=0.$

The following statement lists all integrable Hamiltonians up to canonical transformations (see Section 2).

{\theorem Suppose that non-linear equation {\rm (\ref{Ham})} has infinite hierarchy of higher symmetries 
$$
u_{\tau_k}=F_k(x,u,u_x,\dots),\ \ \ \ k=1,2,\dots.
$$
Then the Hamiltonian $H$ is canonically equivalent to one of the following functions:
\begin{align}
\label{e-a0}
&H= -\frac{u_x^2}{2\,Q(u)^3}+\frac{P(u)}{Q(u)},    \\[4mm]
\label{e-a2-2}
& H=-\frac{u_x^2}{2\,u^3 }+\frac13 P(x) u^3,\\[3mm]
&    H=\sqrt{u_x+P(u)}. \label{eq-c1-a1} 
\end{align}
Here  $P$ and $Q$ are  arbitrary polynomials of degrees not greater than 4,\, and 2, correspondingly.} $\square$

In this paper we consider the fifth order integrable Hamiltonian evolution equations of the form 
\begin{equation}\label{Ham0}
u_t=D_x\left(\frac{\delta H(x,u,u_x,u_{xx})}{\delta u}  \right)=D_x\left(\frac{\p H}{\p u}-D_x\frac{\p H}{\p u_x}+D_x^2\frac{\p H}{\p u_{xx}}\right).
\end{equation}

It turns out that there exist two types of integrable Hamiltonians:
$$
H=a_1 u_{xx}^2+a_2 u_{xx}+a_3 \qquad {\rm and} \qquad H=k_1 u_{xx}+k_2+\sqrt[3]{u_{xx}+k_3 }, 
$$
where $a_i$ and $k_i$ are some functions of $u, u_x$ and $x$.

Using the symmetry approach to integrability \cite{ss,mss},  we obtain a complete list of canonical forms for integrable Hamiltonians $H$. Our proof of the classification statement contains an 
algorithm which allows to bring any integrable Hamiltonian to one of the canonical forms by canonical transformations. 

The list contains 
Hamiltonians for all fifth order commuting flows of equations from Theorem 1. For example, the  Hamiltonian
$$
H_1=\frac{1}{2}u_{xx}^2-5\,uu_x^2+\frac{5}{2}u^4
$$ 
generates the fifth order equation from the KdV hierarchy. Besides these Hamiltonians, we found the following integrable Hamiltonians of second order:
\begin{align}
&H=\sqrt[3]{u_{xx}+(5\,u^2+9\,k)u_x+2\,u\,(u^2+k)(u^2+9\,k)},  \label{Ham1}\\[2mm]
&H=\sqrt[3]{\vphantom{u^2}u_{xx}}+\frac13\,c_0\,u^3+\frac12\,(c_1x+c_2)u^2+(c_3x^2+c_4x)u, \label{Ham2}\\[2mm]
&H=\frac{1}{2}\,{u_{xx}}^2-\frac{5}{6}\,u_x^3+\frac{5}{2}\,u^2u_x^2+\frac{1}{6}\,u^6, \label{Ham3}\\[2mm]
&H=\frac12\,{\frac {{{u_{xx}}}^{2}}{{{u}}^{10}}}-{\frac {20}{3}}\,{\frac {{{u_x}}^{4}}{{{u}}^{12}}}+\frac53\,{\frac {{{u_x}}^{3}}{
{{u}}^{10}}}+\frac52\,{\frac {{{u_x}}^{2} \left( k{u}+2 \right) ^{2}}{{{u}}^{8}}}
+{\frac {20}{3}}\,{\frac {{k}^{3}}{{u}}}+10\,{\frac {{k}^{2}}{{{u}}^{2}}}+8\,{\frac {k}{{{u}}^{3}}}+\frac{8}{3\,u^4}. \label{Ham4}
\end{align} 
The constant $k$ can be reduced to $k=0$ or to $k=1$ by a scaling. The constants $c_i$ are also not essential (see Remark 4). Notice that the equation
$$
u_\tau =D_x\frac{\delta }{\delta u}\left(\frac12\,(c_1x+c_2)u^2+(c_3x^2+c_4x)u\right)
$$
is the classical symmetry for the Hamiltonian equation with $H$ given by (\ref{Ham2}).

It follows from \cite{sw,ow} that if the right hand side of  integrable Hamiltonian equation 
\begin{equation}
u_t=u_n+F(x,u, \, u_x, \, u_{xx}, \dots, u_{n-1}), \qquad u_i=\frac{\partial^i u}{\partial x^i}. \label{scalar}
\end{equation} 
is polynomial and homogeneous, then  the hierarchy of this  equation contains an equation of third or fifth order. This statement looks very credible also 
without any additional restrictions for the right hand side of the equation. The proof in the general case is absent and this statement has a status of the 
conjecture well-known for experts. No counterexamples to this conjecture are known.

Up to this conjecture, in this paper taken together with \cite{meshsok}  we have described simplest equations for all hierarchies of the integrable Hamiltonian equations of the form (\ref{Hamgen}). In other words, any integrable equation (\ref{Hamgen}) is equivalent to a generalized symmetry of one of the equations presented in our papers.  

\section{Preliminaries.}

\subsection{Canonical transformations.} 
It is easy to verify that the equation (\ref{Ham0}) is stable under transformations
$H\to H+D_x f(x,u,u_x)+\lambda u,$ where  the function $f$ and the constant $\lambda$ are arbitrary.

Some transformations preserve the Hamiltonian form  (\ref{Ham0}) changing the right hand side of equation. We call such transformations {\it canonical}.

Consider point transformations of the form
\begin{equation}\label{ptr}
x=\phi(y,v), \qquad  u=\psi (y,v).
\end{equation}
The invertibility of the transformation is equivalent to the inequality  $\Delta =\phi_y\psi_v-\phi_v\psi_y\ne0$. 
Transformation (\ref{ptr}) is called {\it canonical} if $\Delta =\phi_y\psi_v-\phi_v\psi_y=1$.
It is easy to verify that canonical transformations preserve the form of equation (\ref{Ham0}). 
The Hamiltonian of the resulting equation is given by
\begin{equation}\label{trH}
\t H(y,v,v_y)=H\left(\phi(y,v),\psi (y,v),\frac{D_y(\psi)}{D_y(\phi)}\right)D_y(\phi).
\end{equation}

{\bf Example.} Linear   transformations of the form
$$
 x=f(y),\qquad u=\frac{v}{f'}+g(y),\qquad \t H=Hf'.
$$
are canonical for arbitrary functions $f$ and $g$.

\rem If we consider only Hamiltonians that do not depend on $x$ explicitly, we still have non-trivial canonical transformations
$$ 
\begin{array}{c}
x=f(v)+ y,\quad u=v,\qquad \t H=(f'v_y+1)\,H.   \\[4mm]
x=v,\quad  u=f(v)+y, \qquad \t H=H\,v_y.   \qquad \square
\end{array}
$$

Besides (\ref{ptr}) we use the following canonical transformations of a more general form: 

1. Dilatations of the form
$$
 t=\alpha \t  t,\qquad x=\beta y,\qquad u=\gamma v, \qquad \t H =\frac{\alpha }{\beta \gamma^2}H(\beta y,\gamma v)
$$
are admissible for any $H$.
 
2. If $H$ does not depend on $x$,  then the Galilean transformation 
$$
y= x +c\,t,\qquad  v=u, \qquad   \t H=H-\frac{1}{2}\,c\,v^2;
$$
is admissible.

3. If $H=h(u_x)+c\,xu$, where $c$ is a constant, then the following transformation
$$
u\to u+c\,t, \quad H\to H-c\,xu
$$
is admissible.

4.  If $H=h(u_{xx})+(c_1x^2+c_2x)u$, where $c_1$ and $c_2$ are some constants, then the following transformation
$$
u\to u+2\,c_1x\,t,+c_2t, \quad H\to H-(c_1x^2+c_2x)u
$$
is admissible.

\rem Sometimes we use the point transformations (\ref{ptr}) without the condition $\Delta =1$. Using the properties of the Euler operators that are presented in \cite{olv} one can find the form of the transformed
equation (\ref{Ham0}):
\begin{equation}\label{modH}
v_t=f\,D_y\left(f\,\frac{\delta\t H}{\delta v}  \right),
\end{equation}
where $f=\Delta^{-1}=f(y,v)$, $\t H=H\,(D_y\phi)$. One can easily verify the operator $f\circ D_y\circ f$ is the Hamiltonian one for any $f$.

\subsection{Canonical densities.} The necessary integrability conditions for equations (\ref{scalar}) have the form of  an infinite chain of the canonical conservation laws
\begin{equation}
\label{law1}
\frac{d}{dt}(\rho_{n})=\frac{d}{dx}(\theta_{n}), \quad n=-1,0,1,\dots ,
\end{equation}
where the functions $\rho_i$ are defined by a recurrent formula. Following  \cite{sokmesh}, we find that in the case of fifth order equations 
\begin{equation}\label{5ord_eq}
u_t=F(x,u,u_x,u_{2},\dots,u_{5})
\end{equation}
the recurrence can be written as follows
\begin{equation}\label{recc5}
\begin{aligned}
\rho_{n+4}&=\frac{1}{5}\rho _{-1}\theta_n- \frac{1}{5}\rho _{-1}\left(F_0\delta_{n,0}+F_1\rho_n+F_2D_x(\rho_n)+F_2\sum_{-1}^{n} \rho_i\rho_j+F_3D_x^2(\rho_n)\right)\\
&- \frac{1}{5}\rho _{-1}F_3\left(\frac{3}{2}D_x\sum_{-1}^{n} \rho _i\rho _j+\sum_{-1}^{n} \rho _i\rho _j\rho _k\right)- \frac{1}{5}\rho _{-1}F_4\left(D_x^3(\rho _n)+2D_x^2\sum_{-1}^{n} \rho _i\rho _j\right)\\
&- \frac{1}{5}\rho _{-1}F_4\left(-\sum_{-1}^{n}D_x( \rho _i)D_x(\rho _j)+2D_x\sum_{-1}^{n} \rho _i\rho _j\rho _k+\sum_{-1}^{n} \rho _i\rho _j\rho _k\rho _l\right)\\
&-(\rho_{-1})^{-4}\left( \frac{1}{5}D_x^4(\rho_n)+\frac{1}{2}D_x^3\sum_{-1}^{n} \rho_i\rho_j- \frac{1}{2}D_x\sum_{-1}^{n} D_x( \rho _i)D_x(\rho _j)+\frac{2}{3}D_x^2\sum_{-1}^{n} \rho_i\rho_j\rho_k 
\right.\\
&\left.\qquad - \sum_{-1}^{n} \rho_iD_x( \rho _j)D_x(\rho_k)+\frac{1}{2}D_x\sum_{-1}^{n} \rho_i\rho_j\rho_k\rho_l
 +\frac{1}{5}\sum_{-1}^{n}  \rho _i\rho _j\rho _k\rho _l\rho_m-  \rho _{-1}^4\rho_{n+4} \vphantom{\sum_{-1}^{n}}    \right),
\end{aligned}
\end{equation}
where $n=-4,-3,-2,\dots,\ F_n=\p F/\p u_n$ and
$$
\rho_{-1}=F_5^{-1/5}.
$$
 By definition,
$$
\sum_{a}^{b} \rho_{I_1}\cdots \rho_{I_k}=\sum_{\substack{I_s\ge a,\,1\le s\le k,\\ I_1+\cdots +I_k=b  }} \rho_{I_1}\cdots \rho_{I_k}.
$$
In particular, we have
$$
\begin{aligned}
&\sum_{-1}^{-2} \rho_i\rho_j=\rho_{-1}^2,\qquad \sum_{-1}^{-1} \rho_i\rho_j=2\,\rho_{-1}\rho_0,\qquad \sum_{0}^{-1} \rho_i\rho_j\rho_k=0,\\
&\sum_{-1}^{n} \rho_i\rho_j\rho_k=3\rho_{-1}^2\rho_{n+2}+6\rho_{-1}\rho_{0}\rho_{n+1}+\dots,\qquad \sum_{-1}^{n}  \rho _i\rho _j\rho _k\rho _l\rho_m=5 \rho _{-1}^4\rho_{n+4}+\dots
\end{aligned}
$$

The first three canonical densities are given by  
$$
\rho_{-1}=F_5^{-1/5},\qquad \rho_0=-\frac{1}{5}F_4\rho_{-1}^5-2D_x\ln \rho_{-1},
$$$$
\rho_{1}=\frac{1}{2}F_4D_x(\rho_{-1}^4)+\frac{2}{5}\rho_{-1}^4D_xF_4+\frac{2}{25}\rho_{-1}^9F_4^2-\frac{1}{5}\rho_{-1}^4F_3- 3(\rho_{-1})^{-3}(D_x\rho_{-1})^2+2(\rho_{-1})^{-2}D_x^2(\rho_{-1}).
$$

For equations (\ref{Ham0}) we have
\begin{equation}\label{h-5-1}
\rho_{-1}=\left(\frac{\p^2 H}{\p u_{xx}^2}\right)^{-1/5}.
\end{equation}
Let us denote  $\rho_{-1}=a(x,u,u_x,u_{xx}).$ Then
\begin{equation}\label{1st-rho}
\frac{\p^2 H}{\p u_{xx}^2}=a^{-5}.
\end{equation}

We use the following notations: $f\sim g$ iff $f-g\in\I D_x$. In particular, $f\sim0$ iff $f\in\I D_x$. We write here $D_x$ instead of $\frac{d}{dx}$ and $D_t$ instead of $\frac{d}{dt}$  for brevity.
Thus, all  integrability conditions may be written in the following form $D_t\rho_n\sim0,\,n=-1,0,\dots$.

We used the following algorithm of checking whether \label{D-1} a given function $S(x,u, u_{1}, \dots , u_{n})$   belongs to $\I D_x$. At first, $S$ has to be linear in the highest derivative $u_{n}$. 
If it holds true, then  we can subtract from $S$ a function of the form $D_x Q(x,u, u_{1}, \dots , u_{n-1})$ such that the difference does nor depend on $u_n$. Repeating this order lowering procedure, 
we either arrive at the situation when the function is nonlinear in its highest derivative, or we get zero.
 
\section{Classification}

{\theorem If non-linear equation of the form {\rm (\ref{Ham0})} has infinitely many generalized symmetries  
$$
u_{\tau_k}=F_k(x,u,u_x,\dots),\qquad k=1,2,\dots
$$
and has no symmetries of order  $p$, where $1<p<5$, then its Hamiltonian can be reduced to one of the   canonical forms} (\ref{Ham1})--(\ref{Ham4}).
 
\rem If  we have $k\ne0$ in (\ref{Ham1}),  then the scaling $u\to u\,k^{1/2},\ x\to x\,k^{-1},\ t\to t\,k^{-4/3}$ brings $k$ to 1. If $k\ne0$ in (\ref{Ham4}) then we can  reduce $k$ to 1 by the scaling $u\to u\,k^{-1},\ x\to x\,k,\ t\to t\,k^{-4}$.
Therefore we may assume that $k=0$ or $k=1$ for both these Hamiltonians.

\rem All constants in Hamiltonian (\ref{Ham2}) can be normalized by canonical transfor\-mations to $1$ or $0$.
If $c_0\ne0$ then using the dilatation $u\to\alpha u,\ t\to\alpha ^{-1} t,\ \alpha =c_0^{-3/8},$ we obtain $c_0=1$. After that we can apply the transformation $u\to u-\frac12(c_1x+c_2)$ to obtain
the Hamiltonian
$$
H=\sqrt[3]{\vphantom{u^2}u_{xx}}+\frac13\,u^3+(k_1x^2+k_2x)u,
$$
where $k_1$ and $k_2$ are some constants. If $k_1\ne0$ the scaling $u\to u\,\mu,\ x\to x\mu^{-4},\ t\to t\mu ^{-5}$ with $\mu=k_1^{1/10}$ leads to $k_1=1$. Now we reduce $k_2$ to zero by the translation $x\to x-k_2/2$. 
If $k_1=0$ and $k_2\ne0$, we normalize $k_2$ by 1 with the help of a scaling.  Thus, if $c_0\ne0$  we have the following three non-equivalent Hamiltonians:
\begin{align}
&H_1=\sqrt[3]{\vphantom{u^2}u_{xx}}+\frac13\,u^3+x^2u, \label{h-1}\\
&H_2=\sqrt[3]{\vphantom{u^2}u_{xx}}+\frac13\,u^3+x\,u,  \label{h-2}\\
&H_3=\sqrt[3]{\vphantom{u^2}u_{xx}}+\frac13\,u^3.  \label{h-3}
\end{align}
If $c_0=0$ and $c_1\ne0$ then (\ref{Ham2})  is equivalent to 
\begin{equation}\label{h-4}
H_4=\sqrt[3]{\vphantom{u^2}u_{xx}}+\frac12\,x\,u^2.
\end{equation}
In the case $c_0=c_1=0$ the Hamiltonian is equivalent to  
\begin{equation}\label{h-5-2}
H_5=\sqrt[3]{\vphantom{u^2}u_{xx}}. \qquad \square
\end{equation}

\medskip

{\bf Proof of Theorem 2.}  Consider the first integrability condition $D_t(a)\sim 0.$ It can be verified that 
\begin{equation}\label{usl1}
D_t a\sim -\frac{5}{2}\,u_4^2a^{-10}D_x\left(a^5\frac{\p^2a}{\p u_2^2}\right)+ O(3),
\end{equation}
where  the symbol $O(n)$ denotes terms of differential order not greater then $n$. Since the right hand side of (\ref{usl1}) has to belong to $\I D_x$ it 
should be linear in the highest derivative. Therefore we get   
$$
\frac{d}{dx}\left(a^5\frac{\p^2 a}{\p u_{2}^2}\right)=0.
$$
Integrating this equation we obtain that
\begin{equation}\label{usl1-1}
 \frac{\p^2 a}{\p u_{2}^2}=-2\,c_0^2a^{-5},
\end{equation}
where $c_0$ is a constant.  
 
Now we consider  the third  integrability condition and find that 
\begin{align}\label{usl3}
D_t \rho_1 &\sim-10u_5^2a^{-13}\frac{\p a}{\p u_2}\left[2\,u_3\left(a^2\frac{\p a}{\p u_2}-c_0\right)\left(a^2\frac{\p a}{\p u_2}+c_0\right)+a^3D_x\left(a^2\frac{\p a}{\p u_2}\right)\right] \\[2mm]
&-\frac{5}{3}u_4^3a^{-18}\left[2\,u_3\left(a^2\frac{\p a}{\p u_2}-c_0\right)\left(a^2\frac{\p a}{\p u_2}+c_0\right)\left(99\,a^4\left(\frac{\p a}{\p u_2}\right)^2+20\,c_0^2\right)+O(2)\right]+\dots\n
\end{align}
Equating to zero the coefficients at $u_5^2u_3$ and $u_4^3u_3,$ we obtain
$$
\frac{\p a}{\p u_{2}}=z a^{-2},
$$
where $z=\pm c_0$. If $z \ne 0$ we reduce $z$ to 1 by the dilatation $\ds u\to \frac{u}{3 z}$ and arrive at the case  
$$
{\bf A:} \qquad  a=(u_{xx}+q(x,u,u_x))^{1/3}.
$$
If $z=0,$ then we get
$$
 {\bf B:} \qquad a=a(x,u,u_x).
$$

{\bf Case A.} Integrating equation (\ref{1st-rho}) we find that the Hamiltonian is equivalent to 
$$
H=f(x,u,u_x)-\frac92\,a.
$$
The first integrability condition yields $\ds \frac{\p^2 f}{\p u_{x}^2}=0.$  Since the Hamiltonian is defined up to functions from  $\I D_x$ we set without loss of generality   
$$
H=h(x,u)-\frac92\,a,\qquad a=\sqrt[3]{u_{xx}+q(x,u,u_x)}.
$$ 
It follows from the third integrability condition that  
$$
q=q_0+q_1 u_x+q_2 u_x^2+q_3 u_x^3,\qquad q_i=q_i(x,u).
$$

Applying (non-canonical) transformation $y=\phi (x,u),\  v=\psi(x,u),$ we reduce to zero $q_2$ and $q_3$. To do this we can take for $\phi$ and $\psi$ any functionally independent solutions of the system of PDEs
\begin{align*}
&\phi_{xx}=2\,\frac {\phi_{ux}\phi_x}{\phi_u}-\frac {\phi_ x^{2}\phi_{uu}}{\phi_u^{2}}+{q_0}\,\phi_u-{q_1}\,\phi_x
+q_2\,\frac {\phi_x^{2}}{\phi_u}- {q_3}\frac {\phi_x^3}{\phi_u^2},\\
&\psi_{xx}=2\frac{\psi_{xu}\phi_x}{\phi_u}-\frac{\phi_x^2\psi_{uu}}{\phi_u^2}+\frac{2\,h}{\phi_u^3}\left(\phi_{uu}\phi_x-\phi_{ux}\phi_u\right)+q_0\psi_u-q_1\psi_x\\
&\qquad+q_2\,\frac{\phi_x}{\phi_u^2}\left(\phi_u\psi_x-\Delta \right)+q_3\,\frac{\phi_x^2}{\phi_u^3}\left(2\Delta -\phi_u\psi_x\right),
\end{align*}
where $\Delta =\phi_x \psi_u- \phi_u \psi_x$.  

The resulting Hamiltonian has the following structure:  
$$
\tilde H=\tilde h(x,u)-\frac92\,\sqrt[3]{f(x,u)\,u_{xx}+\tilde q_0(x,u)+\tilde q_1(x,u)u_x},
$$
where $f=\Delta^{-1}$. The Hamiltonian form of the corresponding equation (\ref{5ord_eq})  is given by (\ref{modH}). The first canonical density for this equation has the form 
$$
\rho_{-1}=\sqrt[3]{f\,u_{xx}+\tilde q_0+\tilde q_1 u_x}\,f^{-4/5}.
$$
The integrability condition  $D_t(\rho_{-1})\sim 0$ leads to $f=c,$ where $c$ is a non-zero constant. We reduce $c$ to 1 by the transformation $t\to t\,c^{-7/3}.$ This means that for integrable equations the above transformation $y=\phi,\,  v=\psi$ is a canonical one. 
Thus we have shown that the Hamiltonian can be reduced to 
$$
H=h(x,u)-\frac92\,a,\qquad a=\sqrt[3]{u_{xx}+q_0+q_1 u_x},
$$ 
where the Hamiltonian form of the corresponding equation (\ref{5ord_eq})  is given by (\ref{Ham0}).
 
It follows from the third integrability condition that  
$$
q_1=s_2(x)u^2+s_1(x)u+s_0(x).
$$
Under canonical transformations of the form  $y=f(x),\,v=u/f'+g(x)$ this function changes as follows:   $q_1 \to \t q_1=\t s_2 u^2+\t s_1 u+\t s_0$, where
$$
\t s_2= s_2 f',\ \ \ \t s_1=s_1-2\,g\,s_2 f',\ \ \ \t s_0= (f')^{-2}(s_0 f'+f''-g s_1 {f'}^2+g^2 s_2{f'}^3).
$$ 
If $s_2\ne0$, we take $f=5 \int s_2^{-1}dx$ and $g=s_1/10$ to get  $\t s_2=5,\ \t s_1=0$.  
If  $s_2=0$, then  $\t s_1=s_1$. Choosing $g=0$ and taking for $f$ any non-constant solution of the equation $f''+s_0 f'=0$, we arrive at $\t s_0=0$. So we are to consider the cases
$$
{\bf A.1.}\quad    q_1=5 u^2+s(x), \qquad {\bf A.2.}\quad  q_1=s(x)\,u.
$$

{\bf Case A.1.} It follows from the third integrability condition that $s(x)$ is a constant. We denote it by $9\, k.$
Moreover we find that 
$$
q_0=2\,u\,(u^2+k)(u^2+9\,k).
$$
For such $q_1$ and $q_0$ the third integrability condition turns out to be equivalent to $h=c_2 u^2+c_1 u$ where $c_1$ and $c_2$ are some constants. As it was 
mentioned in Section 2.1, $c_1$ is a trivial constant and the term $c_1u^2$ can be removed
by the Galilean transformation. So, we obtain the integrable Hamiltonian (\ref{Ham1}).

{\bf Case A.2.} From the first and third integrability conditions we find that
$$
H=\frac13p_0(x)\,u^3+\frac12\,p_1(x)u^2+p_2(x)u-\frac{9}{2}\sqrt[3]{u_{xx}+r_1(x)u+r_0(x)},
$$
To simplify this Hamiltonian we use linear (non-canonical) transformations of the form $y=\phi(x),\ v=u\,g(x)+\psi(x)$. Taking a solution of the PDE system
$$
\phi '=g^2, \quad g''=2\,g^{-1}{g'}^2+r_1g,\quad \psi''=2\,g^{-1}{g'}\psi'+r_0g
$$ 
for $g, \phi$ and $\psi,$ we arrive at  the transformed Hamiltonian 
$$
H=F(x,u)- \frac{9}{2}\Big(u_{xx}f(x)\Big)^{1/3},
$$
where  $f=g^{-3}$ and $F=h/g^2$ is the third degree polynomial with respect to $u$. 

Since the transformation was not canonical, the  Hamiltonian equation now takes the form (\ref{modH}). The first canonical density now reads as $\rho_{-1}={u_{xx}}^{1/3}\,f^{-7/15}$.
The first integrability condition immediately leads to the fact that $f$ is a constant. By a dilatation of the form $t\to \lambda t$ we reduce $f$ to 1. This means that for any integrable 
Hamiltonian we can reduce the coefficients $r_1, r_0$ to zero by a canonical transformation. Now we find that the polynomial $F$ has the form
$$
F=\frac13\,{c_0}\,u^{3}+\frac12\,( {c_1}\,x+{c_2})u^{2}+( {c_3}\,{x}^{2}+{c_4}\,x)u,
$$
where $c_i$ are some constants. Thus, we obtain Hamiltonian (\ref{Ham2}). All constants $c_i$ are inessential (see Remark 4).


{\bf Case B.} Integrating equation (\ref{1st-rho}), we find that without loss of generality the Hamiltonian can be written in the form   
\begin{equation}\label{h-b}
H=h(x,u,u_x)+\frac{u_{xx}^2}{2\,a^5}.
\end{equation}
It follows from the first integrability condition that 
$$
D_x\left(a^3\frac{\p^2a}{\p u_x^2}\right)=0.
$$
The solution of this equation is given by  
$$
a=\sqrt{a_0+a_1u_x+a_2u_x^2}\,,\qquad a_1^2-4\,a_2a_0=c^2,
$$
where $a_i=a_i(x,u)$, and  $c$ is a constant. It can be verified (see \cite{meshsok}) that $a_2$ can be reduced to zero by a canonical transformation $y=\phi(x,u),\ v=\psi(x,u)$. 
Now the condition $a_1^2-4\,a_2a_0=c^2$ implies that $a_1$ is a constant. 
Using a scaling of the form  $u\to \lambda u,$ we arrive at the following two subcases:  
$$
{\bf B.1.}\ a=a(x,u),\qquad \qquad  {\bf B.2.}\ a=\sqrt{u_x+q(x,u)}. 
$$

{\bf Case B.1.} It follows from the first integrability condition that  
$$
\frac{\p^3a}{\p u^3}=0,\qquad \frac{\p}{\p x}\left(\frac{\p^2a^2}{\p u^2}-3\left(\frac{\p a}{\p u}\right)^2\right)=0,
$$
which means that
$$
a=s_0+s_1u+s_2u^2,  \quad \qquad  \delta =s_1^2-4\,s_2s_0=constant, 
$$
where $s_i=s_i(x).$ Using linear canonical transformations
$$
y=\phi (x),\qquad  v=u(\phi ')^{-1}+f(x) 
$$
and scalings, we can simplify the function $a.$ As a result we arrive at the subcases 
$$
{\bf B.1.1.}\ a=1,\qquad {\bf B.1.2.}\ a=u,\qquad {\bf B.1.3.}\ a=u^2+z,
$$
where $z$ is a constant. 

{\bf Case B.1.1.}  It follows from the third integrability condition that  
$$
H=\frac12u_{xx}^2+ \frac13f_1u_x^3+\frac12f_2u_x^2+f_3, \qquad f_i=f_i(x,u).
$$ 
The fifth integrability condition implies   $f_1=c_0,\ f_2=c_1u^2+a_2u+a_3$, where $a_i=a_i(x)$ and $c_i$ are constants such that   $c_0(4\,c_0^2-5\,c_1)=0$.

{\bf Subcase B.1.1.a.} If $c_0\ne0$, then $c_1=\frac45\,c_0^2\ne 0$. Using a canonical transformation of the form  $u\to u-f(x)$, we reduce $a_2$ to zero. It follows from the fifth integrability condition that  
$$
H=\frac12\,{u_{xx}}^{2}+\frac13\,{{u_x}}^{3}{c_0}+\frac25\,{{u_x}}^{2}{{c_0}}^{2}{{u}}^{2}+{\frac {8}{1875}}\,{{u}}^{6}{{ c_0}}^{4}.
$$
Using a scaling of the form $u\to\lambda u,$ we put  $c_0=-5/2$ getting Hamiltonian (\ref{Ham3}).

{\bf Subcase B.1.1.b.} If  $c_0=0$ and $c_1\ne0$, we reduce the function $a_2$ to zero by the canonical transformation $u\to u-a_2/(2c_1)$. Assuming  $a_2=0$, we find from the fifth integrability condition that   
$$
H=\frac12u_{xx}^2+ \frac12(c_1u^2+c_2)u_x^2+\frac{c_1^2}{100}u^6+\frac{c_1c_2}{20}u^4+\frac12c_3u^2+c_4u.
$$
The constant $c_4$ is trivial (see Section 2.1) and $c_3$ can be annihilated by the Galilean transfor\-mation. 
By a transformation of the form $u\to \lambda u$ we reduce $c_1$ to 10 and get 
$$
 H=\frac{1}{2}\,u_{xx}^2+5\,u^2u_x^2+u^6+\frac{1}{2}\,c(u_x^2+u^4).  
$$
The corresponding evolution equation
$$
u_t=D_x\left(u_4-10\,u(u\,u_{2}+u_x^2)+6\,u^5+c\,(2\,u^3-u_{xx})\right)  
$$
is a higher symmetry of the mKdV equation $u_t=D_x(u_{xx}-2\,u^3)$. This equation has Hamiltonian (\ref{e-a0}) with $Q=1$ and $P=-\frac12u^4$.

{\bf Subcase B.1.1.c.} If  $c_0=c_1=0$, integrability conditions imply $a_2=k=constant$ and
$$
H=\frac12u_{xx}^2+\frac12(k\,u+a_3)u_x^2+\frac{40\,c_3+3\,k^2}{40}u^4+\frac13a_4u^3+\frac12a_5u^2+a_6u, \qquad a_i=a_i(x),
$$ 
where $k(k^2+20\,c_3)=0$. 

{\bf I.} If  $k\ne0$, then $c_3=-k^2/20$. Using the shift $u\to u-a_3/k,$ we reduce $a_3$ to zero. After that integrability conditions lead to the fact that $a_5$ and $a_6$ 
are constants and that  $a_4=0$. The constant $a_6$ is trivial  and $a_5$ can be reduced to 0 by the Galilean transformation. 
Using the transformation $u\to-(10/k)\,u$, we obtain
$$
 H=\frac{1}{2}\,u_2^2-5\,uu_x^2+\frac52\,u^4,  \qquad u_t=D_x\left(u_4+10\,u(u_{2}+u^2)+5\,u_x^2\right).
$$
The latter equation belongs to the hierarchy of the KdV equation $u_t=D_x(u_2+3\,u^2)$.  This equation has Hamiltonian (\ref{e-a0}) with $Q=1$ and $P=u^3$.

{\bf II.} If  $k=0$, then the Hamiltonian 
$$
H=\frac12u_{xx}^2+\frac12a_3u_x^2+c_3u^4+\frac13a_4u^3+\frac12a_5u^2+a_6u, \qquad a_i=a_i(x)
$$ 
can be simplified by a canonical transformation of the form $u\to u-f(x)$. The coefficients of the Hamiltonian are transformed as follows:  
$$
\begin{aligned}
&\t a_3=a_3,\quad \t c_3=c_3,\qquad \t a_4=a_4-12\,c_3f,\qquad \t a_5=a_5+12\,c_3f^2-2\,a_4f,\\
& \t a_6=a_6-f^{(4)}-4\,c_3f^3+a_3f^2-a_5f+(f'a_6)'.
\end{aligned}
$$
Consider the following alternatives:
\begin{itemize}
\item[a)] If $c_3\ne0,$  then we take $f=a_4/(12\,c_3)$ to get $\t a_4=0.$
\item[b)] If  $c_3=0,\ a_4\ne0,$ then choosing $f=a_5/(2\,a_4)$  we get  $\t a_5=0.$
\item[c)] If  $c_3=0,\ a_4=0,$ then we consider the normalization $\t a_6=0$.
\end{itemize}
In the cases a) and b) we obtain non-integrable equations  $u_t=D_x(u_4+4c_3u^3+c\,u_2)$ and $u_t=D_x(u_4+c_4u^2+c\,u_2),$ $c_4\ne0$, correspondingly.  
For the first of them the ninth integrability condition is not fulfilled and for the second the condition number eleven is broken.  

In the case c)  a linear equation appears with the Hamiltonian $\ds H=\frac{1}{2}(u_{xx}^2+a(x)u_x^2+b(x)u^2)$.

{\bf Case B.1.2.} Here we consider the Hamiltonians of the form
$$
H=h(x,u,u_x)+\frac{u_{xx}^2}{2\,u^5}.
$$
It follows from the third integrability condition that  
$$
D_x\left(u^7\frac{\p^4h}{\p u_x^4}\right)=0\ \text{or }\ h=c_1\frac{u_x^4}{u^7} +q_1u_x^3+q_2u_x^2+q_3u_x+q_4,\qquad q_i=q_i(x,u).
$$
The dependence of the functions $q_i$ on $u$ can be specified by the third and fifth integrability conditions:
\begin{align}\label{Hb1-2}
H=\frac{u_{xx}^2}{2\,u^5}-\frac{15}{8}\frac{u_x^4}{u^7}+\frac{1}{2}u_x^2\left(\frac{s_1}{u^5}+\frac{s_2}{u}\right)+\frac{s_2^2}{50}u^5- \frac{1}{2}s_3u^2-\frac{u^{-3}}{150}(10\,s_1''-3\,s_1^2)
-\frac{u}{15}(5\,s_2''+s_1s_2),
\end{align}
where  $s_i=s_i(x)$. 
The functions $s_1,\, s_2$ and $s_3$ are changed under the canonical transformation $y=\phi(x),\ v=u/\phi'$ as follows:
$$
\t s_1=\frac12(\phi')^{-4}(10\,\phi'''\phi'-15\,\phi''^2+2\,s_1\phi'^2),\qquad \t s_2=s_2\phi '^2,\qquad \t s_3=s_3\phi'.
$$
Choosing $\phi$ as a nonzero solution of the equation $10\,\phi'''\phi'-15\,\phi''^2+2\,s_1\phi'^2=0,$ we obtain $\t s_1=0$. In this case the integrability conditions 1-7 are equaivalent to equations
$$
s_2^{(5)}=0,\qquad s_3'''=0,\qquad s_3s_2'=2s_2s_3'. 
$$
Two solutions: (i) $s_3\ne0,\ s_3=c_0+c_1x+c_2x^2,\ s_2=k\,s_3^2$, and (ii) $s_3=0$ and $s_2(x)$ is a polinomial of degree not greater than 4, lead to
$$
H_1=\frac{u_{xx}^2}{2\,u^5}-\frac{15}{8}\frac{u_x^4}{u^7}+s_3^2\frac{k\,u_1^2}{2\,u}+\frac{k^2}{50}s_3^4u^5- \frac{1}{2}s_3u^2-\frac{2}{3}k\,u(s_3s_3''+{s_3'}^2),
$$
and
$$
H_2=\frac{u_{xx}^2}{2\,u^5}-\frac{15}{8}\frac{u_x^4}{u^7}+\frac{s_2}{2}\frac{u_x^2}{u}+\frac{s_2^2}{50}u^5-\frac{u}{3}s_2''.
$$
Equations (\ref{Ham0}) corresponding to both $H_1$ and $H_2$ have third order symmetries of the form
$$
u_\tau=D_x\left(\frac{u_{xx}}{u^3}-\frac{3}{2}\frac{u_x^2}{u^4}+u^2P(x)\right),
$$
where  $P=-\frac{3}{5}ks_3^2$ and  $P=-\frac{3}{5}s_2$ correspondingly. These symmetries are generated by Hamiltonians of the form (\ref{e-a2-2}).

{\bf Case B.1.3.} For Hamiltonians of the form
\begin{equation}\label{B13}
H=h(x,u,u_x)+\frac{u_{xx}^2}{2\,\mu ^5}, \qquad \mu =u^2+z
\end{equation}
the first integrability condition leads to 
$$
\frac{\p h}{\p x}=D_x f(x,u)
$$
for some function $f$. Therefore $h$ has the form $h=D_x g+h_0(u,u_x)$ and without loss of generality we assume that $h=h(u,u_x)$.

It follows from the third integrability condition that  
$$
D_x\left(\mu ^7\frac{\p^4h}{\p u_x^4}+160\,\mu \right)=0,
$$
and therefore the Hamiltonian is equivalent to 
\begin{equation}\label{Hb1-3}
H=\frac{u_{xx}^2}{2\,\mu ^5}-\frac{20}{3}\frac{u_x^4}{\mu^7}\big(c_1+\mu\big)+\frac13q_1u_x^3+\frac12q_2u_x^2+q_3,
\end{equation}
where $q_i=q_i(u)$ and $c_1$ is a constant. From the third and fifth integrability conditions we derive 
two relations  $(8\,c_1+9\,z)(4\,c_1+5\,z)=0$ and $(8\,c_1+9\,z)(6\,c_1+7\,z)=0$, which implies $c_1=-9\,z/8$. Taking this into account, we find 
that the third integrability condition is equivalent to the following relations:
\begin{align}
&\mu^3q_2'''+24\,u\,\mu^2q_2''+12\,\mu(14\,\mu-11\,z)q_2'+24\,u(14\,\mu-5\,z)q_2=0, \label{eq_b13-1}\\
 &z\,(5\,q_3'''-2\,\mu^5q_2q_2'-18\,u\,\mu^4q_2^2)=0, \qquad z\,q_1=0.\label{eq_b13-2}
\end{align}
From the fifth integrability condition we additionally find
\begin{equation}\label{eq_b13-3}
u \,q_1'=-10\,q_1,\qquad q_1(5\,u^2q_2''+70\,uq_2'+210\,q_2-8\,u^{12}q_1^2)=0.
\end{equation}

Consider the following two subcases corresponding to $z=0$ and $z\ne0$ in (\ref{Hb1-3})--(\ref{eq_b13-2}). 

{\bf Subcase B.1.3.a.} If $ z=0$, we find from (\ref{eq_b13-1}) and (\ref{eq_b13-3}) that
$$
q_1=k_1u^{-10},\qquad q_2=k_2u^{-8}+k_3u^{-7} +k_4u^{-6},
$$
where $k_i$ are constants.  The fifth integrability condition leads to  
$$
q_3=\frac12 k_5u^{-2}-k_6u^{-1}+\frac{4\,k_1^2-15k_2}{1500\,u^4}(k_2+3\,k_3u),
$$
where 
$$
k_1 (5\,k_2-4\,k_1^2)=0,\quad (5\,k_2-4\,k_1^2)(10\,k_6+3\,k_3k_4)=0,\quad (5\,k_2-4\,k_1^2)(5\,k_3^2+2\,k_0k_4-100\,k_5)=0. 
$$
Additional algebraic relations for $k_i$ follows from the seventh integrability condition. The system of algebraic equations thus obtained has four solutions. 
One solution corresponds to the integrable Hamiltonian (\ref{Ham4}).  The remaining solutions generate the following Hamiltonians:
$$
\begin{aligned}
H_1&=\frac12\,{\frac {{{u_{xx}}}^{2}}{{{u}}^{10}}}-{\frac {20}{3}}\,{\frac {{{u_x}}^{4}}{{{u}}^{12}}}
+{\frac {{{u_x}}^{2}}{{{u}}^{8}}}({c_1}\,{{u}}^{2}+2\,{c_2}\,{u}+{c_3})+\frac{{c_3}}{25}\,{\frac {5\,{c_1}\,{{u}}^{2}+6\,{c_2}\,{u}+{c_3} }{{{u}}^{4}}}+\frac25\,{c_2}\,{\frac {2\,{c_1}\,{u}+{c_2}}{{{u}}^{2}}},\\
H_2&=\frac12\,{\frac {{{u_{xx}}}^{2}}{{{u}}^{10}}}-{\frac {20}{3}}\,{\frac {{{u_x}}^{4}}{{{u}}^{12}}}
+\frac12\,\frac{{u_x}^{2}}{{{u}}^{7}}( 10\,{c_1}+{c_2}\,{u})+\frac52\,{\frac {{{c_1}}^{2}}{{{u}}^{2}}}+{\frac {{c_2}\,{c_1}}{{u}}},\\
H_3&=\frac12\,{\frac {{{u_{xx}}}^{2}}{{{u}}^{10}}}-{\frac {20}{3}}\,{\frac {{{u_x}}^{4}}{{{u}}^{12}}}+\frac{c_1}{2}\frac{u_x^2}{u^6}-\frac{c_2}{u}.
\end{aligned}
$$
The Hamiltonians $H_1$ and $H_2$ correspond to fifth order symmetries for equations  
$$
u_t=D_x\left(\frac{u_{xx}}{u^6}-3\frac{u_x^2}{u^7}+\frac15(2\,c_2u^{-2}+c_3u^{-3})\right)
$$
and
$$
u_t=D_x\left(\frac{u_{xx}}{u^6}-3\frac{u_x^2}{u^7}+\frac{c_1}{u^{2}}\right),
$$
correspondingly. The first of  these equations has Hamiltonian (\ref{e-a0}) with $Q=u^2$ and $\forall P$. The second  equation has Hamiltonian (\ref{e-a0}) with $Q=u^2$ and $P=-c_1u$.

If $c_2\ne0,$ the equation corresponding to $H_3$ does not satisfy the eleventh integrability condition.  In the case  $c_2=0$ we get a symmetry of equation  
$$
u_t=D_x\left(\frac{u_{xx}}{u^6}-3\frac{u_x^2}{u^7}\right).
$$
 This equation has Hamiltonian (\ref{e-a0}) with $Q=u^2$ and $P=0$. The reciprocal  transformation $dy=\rho_{-1}dx+\theta_{-1}dt,\ v(t,y)=1/u(t,x)$ linearizes the equation. The Hamiltonian equation 
for $H_3$ with $c_2=0$ is also  linearizable.

{\bf Subcase B.1.3.b.} If $z\ne0$ in Hamiltonian (\ref{B13}), then we find from equation (\ref{Hb1-3}), (\ref{eq_b13-1}) and (\ref{eq_b13-2}) that
\begin{equation}\label{Hb1-b}
H=\frac{u_{xx}^2}{2\,\mu ^5}-\frac{5}{6}\frac{u_x^4}{\mu ^7}(8\,\mu -9\,z)+5\,\phi\,u_x^2+\psi, 
\end{equation}
where $\phi =q_2/10,\quad \psi =q_3$,
$$
\begin{aligned}
&\mu=u^2+z,\qquad \phi =-{\frac {z\,(2\,{k_1}{u}+{k_2}) }{{\mu}^{5}}}+{\frac {{k_1}{u}+{k_2}}{{\mu}^{4}}}+{\frac {{k_3}}{{\mu}^{3}}},\\
&\psi =\frac{z}{2}\,{\frac {4\,{{k_1}}^{2}z-4\,{k_1}\,{k_2}\,{u}-{{k_2}}^{2}}{{\mu}^{3}}}
+{\frac {3\,{k_1}\,{k_2}\,{u}-4\,{{k_1}}^{2}z+{{k_2}}^{2}}{{\mu}^{2}}}
+\frac52\,{\frac {4\,{k_1}\,{k_3}\,{u}+2\,{k_2}\,{k_3}+{{k_1}}^{2}}{\mu}}.
\end{aligned}
$$
Here $k_i$ are constants. The corresponding Hamiltonian equation is a symmetry of the following third order equation  
$$
u_t=D_x\left(\frac{u_{xx}}{\mu ^3}-3\frac{u\,u_x^2}{\mu ^4}+\frac{k_2u-2\,z\,k_1}{\mu ^2}+\frac{k_1}{\mu }\right).
$$
 This equation has Hamiltonian (\ref{e-a0}) with $Q=\mu=u^2+z,\ z\ne0$ and $\forall P$. 

{\bf Case B.2.} In this case $a=\sqrt{u_x+q(x,u)}$ and the Hamiltonian has the form
$$
H=h(x,u,u_x)+\frac{1}{2}\frac{u_{xx}^2}{a^5}.
$$
The first integrability condition implies  the  following equation
$$
\begin{aligned}
\frac{\p}{\p u_x}\left(a^3\frac{\p^2 h}{\p u_x^2}\right)=\frac{105}{8}f_1\,a^{-8}-\frac{5}{2}\,f_2a^{-6}+\frac{5}{8}\,f_3a^{-4},
\end{aligned}
$$
where
\begin{equation}\label{f1-3}
f_1=(q_x-qq_u)^2,\quad f_2=2\,q^2q_{uu}-4\,qq_{ux}-5\,q_uq_x+5\,qq_u^2+2\,q_{xx},\quad f_3=8\,qq_{uu}-q_u^2-8\,q_{ux}.
\end{equation}
Integrating the equation for $h,$ we found that  the Hamiltonian is equivalent to  

\begin{equation}\label{Hb2}
H=h_1(x,u)+a\,h_2(x,u)-\frac{1}{2}f_1a^{-5}+\frac13f_2a^{-3}-\frac{5}{6}f_3a^{-1}+\frac{1}{2}\frac{u_{xx}^2}{a^5},
\end{equation}
where $f_i$ are defined by (\ref{f1-3}), $h_1$ and $h_2$ are some functions.  It follows  from the first integrability condition that
\begin{align}
&q^{(5)}(u)=0,\qquad h_1'''(u)=0,\qquad D_x\left(h_2-\frac{10}{3}q_{uu}\right)=0, \label{1eq}\\
&q_xh_{1,uu}+2qh_{1,uux}-q_uh_{1,ux}-h_{1,uxx}=0. \label{2eq}
\end{align}
Integrating (\ref{1eq}), we get
$$
q=q_1u^4+q_2u^3+q_3u^2+q_4u+q_5,\quad h_1=\frac12s_1u^2+s_2u,\quad h_2=\frac{10}{3}(q_{uu}+c_0),
$$
where $q_i=q_i(x)$, $s_i=s_i(x)$ and $c_0$ is a constant.

The functions $h_1$ and $q_i$ transform under canonical transformation $y=\phi(x),\ u=v/\phi'+\psi(x)$ as follows: 
\begin{equation}\label{h1}
\t h_1=\frac12s_1\phi' v^2+(s_2-s_1\phi'\psi)v
\end{equation}
and
\begin{equation}\label{tild-q}
\begin{aligned}
&\t q_1=q_1{\phi'}^2,\quad \t q_2={\phi'}(q_2-4\,q_1{\phi'}\psi ),\quad  \t q_3=q_3-3\,q_2{\phi'}\psi+6\,q_1({\phi'}\psi)^2,\\
& \t q_4={\phi'}^{-1}(q_4-2\,q_3{\phi'}\psi+3\,q_2({\phi'}\psi)^2-4\,q_1({\phi'}\psi)^3-{\phi'}^{-1}\phi ''),  \\
& \t q_5={\phi'}^{-2}\big(q_5-q_4{\phi'}\psi+q_3({\phi'}\psi)^2-q_2({\phi'}\psi)^3+q_1({\phi'}\psi)^4-(\phi' \psi)'\big).
\end{aligned} 
\end{equation}

Let us consider the following two alternatives
$$
{\bf B.2.a.}\quad h_1\ne0; \quad \qquad {\bf B.2.b.}\quad h_1=0.
$$

{\bf Subcase B.2.a.1.} Suppose $s_1\ne0$; taking $\phi'=1/s_1$ and $\psi =s_2,$ we obtain $\t h_1=\frac12 v^2$. In this case it follows from equation  (\ref{2eq}) that $q_x=0$. Since the Hamiltonian does not depend on $x$  we may remove the term $\frac12 v^2$ from $\t H$ by the Galilean transformation to reduced $H$ to the following form 
\begin{equation}\label{Hb2-a}
H=\frac{1}{2}\frac{u_{xx}^2}{a^5}+\frac{10}{3}a\,(q''+c_0)-\frac{(qq')^2}{2\,a^5}+\frac{2q^2q''+5\,q{q'}^2}{3\,a^3}+\frac{5}{6}\,a^{-1}(q'^2-8\,qq''),
\end{equation}
where $a=\sqrt{u_x+q(u)}$ and $q^{(5)}=0$.  Equation (\ref{Ham0}) generated by this Hamiltonian is a fifth order symmetry for the equation (see \cite{meshsok})
$$
u_t=D_x\left(\frac{u_{xx}}{{a}^{ 3}}+q'\,\Big(\frac{3}{a}-\frac{q}{a^{3}}\Big)\right).
$$
 This equation has Hamiltonian (\ref{eq-c1-a1}) with  $P=q$. 

{\bf Subcase B.2.a.2.} If $s_1=0$ then $\t h_1=h_1=s_2 u.$ In the case $s_2'=0$ the term  $h_1=c\,u$ is trivial and we have the contradiction 
$h_1=0$, hence    $s_2'\ne0$. Then  equation (\ref{2eq}) takes the form $q_u=-s_2''/s_2'$, therefore $q_1=q_2=q_3=0$ and $s_2''+q_4s_2'=0$. 
It follows from   formulas (\ref{tild-q}) that we can choose   $\phi$ and $\psi$ such that  
$\t q_4=\t q_5=0$. Now we have $q=0$ and therefore  $s_2''=0$.  The Hamiltonian takes the  form
$$
H=c\,x\,u+\frac{10}{3}c_0u_x^{1/2}+\frac{1}{2}\frac{u_{xx}^2}{u_x^{5/2}},
$$
 where $c$ is a constant. The transformation $u\to u+ct$ brings $c$ to zero and the Hamiltonian turns out to be a particular case of (\ref{Hb2-a}).

{\bf Subcase B.2.b.} If $h_1=0$  then  
\begin{equation}\label{Hb2-3}
H=\frac{10}{3}(q_{uu}+c_0)\,a-\frac{1}{2}f_1a^{-5}+\frac13f_2a^{-3}-\frac{5}{6}f_3a^{-1}+\frac{1}{2}\frac{u_{xx}^2}{a^5},
\end{equation}
where $f_i$ are given by (\ref{f1-3}).
It easy to see from  (\ref{tild-q}) that we can reduce the Hamiltonian to one of the following :
$$
\begin{aligned}
&{\bf  B.2.b.1.}\, q_1=1,\ q_2=0;\ \ {\bf  B.2.b.2.}\, q_1=0,\,q_2=1,\,q_3=0;\ \ {\bf  B.2.b.3.}\, q_1=q_2=q_4=q_5=0.
\end{aligned}
$$ 
The first integrability condition results in the equation $q_{x}=0$ for all of these cases. Therefore we have Hamiltonians that can be obtained from formula
 (\ref{Hb2-a})  as partial cases. The theorem is proved.  $\square$

\medskip

\rem  The equation with Hamiltonian (\ref{Ham3}) is well known. It is just equation (3.6) in the list from the survey \cite{sokmesh} . 
The Hamiltonian equation corresponding to (\ref{Ham4}) can be reduced to an equation of the form 
\begin{equation}\label{sep5ord_eq}
u_t=u_5+F(x,u,u_x,u_{2},u_3,u_{4})
\end{equation}
by the standard reciprocal transformation  (see \cite{mss}, section 1.4)
\begin{equation}\label{reci}
dy=\rho_{-1}dx+\theta_{-1}dt, \qquad v(t,y)=u(t,x), 
\end{equation}
where $\rho_{-1}$ is the first canonical density and $\theta_{-1}$ is the correspondent flux. The resulting equation coincides with equation (3.12) from \cite{sokmesh} 
up to the differential substitution $v=2\,w_y^{-1/2}$. Notice that the transformation $v=(w-k/2)^{-1}$ leads to a simpler then (3.12) equation 
$$
w_{{t}}=D_y\Big({w_4}+10\,{w_y}\,{w_{yy}}-20\,{w_{yy}}\,{{w}}^{2}-20\,{{w_y}}^{2}{w}+ (k-2\,{w})^{2}(4\,{{w}}^{3}+4\,k\,{{w}}^{2}+3\,{k}^{2}{w}+2\,{k}^{3})\Big).
$$

To prove the integrability of equations with Hamiltonians (\ref{Ham1}) and (\ref{Ham2}) one could find differential substitutions that reduce them to 
known equations of the form (\ref{sep5ord_eq}). We have verified that these two equations have local conservation laws of 
orders 3 and 5. Also we have found generalized symmetries of order 7 for these equations. In the both cases 
these symmetries have the form
$$
u_\tau =D_x\frac{\delta \rho_1}{\delta u},
$$
where $\rho_1$ is the canonical density for the corresponding fifth order equation. In particular,
for equation with Hamiltonian (\ref{Ham2}) we have $\rho_1\sim u_{xxx}^2u_{xx}^{-7/3}$ and the seventh  order symmetry is given by
$$
u_\tau =D_x\frac{\delta }{\delta u}\left(\frac{u_{xxx}^2}{{u_{xx}}^{7/3}}\right). \qquad   \square
$$

\bigskip

{\bf Acknowledgments.}
The authors would like to thank   B. Dubrovin, V. Kac and O. Mokhov  for useful discussions. 
VS is thankful to IHES for its support and hospitality.


\end{document}